\documentclass[conference,letterpaper]{IEEEtran}
\usepackage{geometry}
\usepackage[utf8]{inputenc}
\usepackage[T1]{fontenc}
\usepackage{url}
\usepackage{ifthen}
\usepackage{cite}
\usepackage{graphicx}
\usepackage{subfigure}

\usepackage{cite}
\usepackage{mathrsfs}
\usepackage{xcolor}
\usepackage{amssymb}
\usepackage{amsthm}
\usepackage{multirow}
\usepackage{threeparttable}
\usepackage{booktabs}
\usepackage{epstopdf}
\usepackage{stmaryrd}
\usepackage{bm}

\newtheorem{theorem}{Theorem}
\newtheoremstyle{noparens}%
{}{}%
{\itshape}{}%
{\bfseries}{.}%
{ }%
{\thmname{#1}\thmnumber{ #2}\mdseries\thmnote{ #3}}
\theoremstyle{noparens}

\newtheorem{lemma}[theorem]{Lemma}

\usepackage[cmex10]{amsmath} 

\begin{document}

\title{On the Convergence Speed of Spatially Coupled LDPC Ensembles Under Window Decoding
}


\author{\IEEEauthorblockN{1\textsuperscript{st} Qingqing Peng}
\IEEEauthorblockA{\textit{School of Mathematics} \\
\textit{Shandong University}\\
Jinan, China \\
Email: pqing@mail.sdu.edu.cn}
\and
\IEEEauthorblockN{2\textsuperscript{nd} Dongxu Chang}
\IEEEauthorblockA{\textit{School of Mathematics} \\
\textit{Shandong University}\\
Jinan, China \\
Email: dongxuchang@mail.sdu.edu.cn}
\and 
\IEEEauthorblockN{3\textsuperscript{rd} Guanghui Wang}
\IEEEauthorblockA{\textit{School of Mathematics} \\
\textit{Shandong University}\\
Jinan, China \\
Email: ghwang@mail.sdu.edu.cn}
\and 
\IEEEauthorblockN{4\textsuperscript{th} Guiying Yan}
\IEEEauthorblockA{\textit{Academy of Mathematics and Systems Science, CAS} \\
\textit{University of Chinese Academy of Sciences}\\
Beijing, China \\
Email: yangy@amss.ac.cn}
}

\maketitle



\begin{abstract}
It is known that windowed decoding (WD) can effectively balance the performance and complexity of spatially coupled low-density parity-check (LDPC) codes. In this study, we show that information can propagate in a wave-like manner at a constant speed under WD. Additionally, we provide an upper bound for the information propagation speed on the binary erasure channel, which can assist in designing the number of iterations required within each window.
\end{abstract}


\section{Introduction}
 Low-density parity-check (LDPC) codes are widely used for their outstanding performance under low-complexity belief propagation (BP) decoding algorithms. However, BP decoders are suboptimal compared to Maximum A Posteriori (MAP) decoders, which are impractical due to their prohibitively high complexity. Many studies have aimed to reduce the performance gap between BP and MAP decoding \cite{sridharan2004convergence,lentmaier2005terminated, lentmaier2010iterative,lentmaier2010asymptotically}. In recent years, spatially coupled LDPC codes, a subclass of convolutional LDPC codes introduced by Felström and Zigangirov \cite{felstrom1999time}, have been shown to significantly improve the BP performance toward the MAP performance of the underlying LDPC codes \cite{lentmaier2005terminated, lentmaier2010iterative,lentmaier2010asymptotically}. This phenomenon is referred to as ``threshold saturation''. Kudekar, Richardson, and Urbanke provided rigorous proof of the threshold saturation phenomenon, demonstrating that the BP threshold of a spatially coupled LDPC ensemble tends to its MAP threshold on any binary symmetric memoryless channel \cite{kudekar2011threshold,kudekar2010threshold,kudekar2013spatially}. Subsequently, the potential function method was introduced to offer a more simple proof
\cite{yedla2012simple, yedla2012simple1,kumar2012proof,kumar2014threshold}.

Moreover, the potential function can also be employed to analyze the convergence speed of the spatially coupled LDPC ensemble under conventional full BP decoding \cite{kudekar2012wave,aref2013convergence}. Kudekar, Aref, and others suggest that when the channel parameter lies between the BP and MAP thresholds, the average error probability computed via density evolution (DE) exhibits a wave-like profile, with lower values at the boundary nodes and higher values in the middle. This wave maintains its shape and propagates toward the middle nodes at a constant speed as the iterations proceed. They also provide bounds on this speed, which help determine the number of iterations required to achieve successful decoding under conventional full BP, and guide the optimization of the degree distribution to maximize propagation speed for a given target channel.

Windowed decoding (WD) is a decoding scheme specifically designed based on the structure of spatially coupled LDPC codes, enabling a balance between decoding complexity and performance compared to full BP decoding \cite{klaiber2018avoiding, zhu2021modeling, schmalen2014spatially, iyengar2011windowed}. The window size $W$ and the number of iterations $T$ within a single window are key parameters in WD, directly impacting the code performance, decoding complexity, and latency. Extensive simulation studies have been conducted to explore how to optimally set \( W \) and \( T \) \cite{schmalen2016design,battaglioni2017design,naseri2019spatially,frenzel2020comparison,battaglioni2017complexity}. However, there are currently no comprehensive theoretical results that specify how to set the number of iterations within a single window to ensure successful decoding.

In this paper, we investigate the convergence speed of WD over spatially coupled LDPC ensembles, which provides insights into how to appropriately configure the number of iterations per window. It is shown that information propagates in a wave-like manner at a constant speed. Specifically, after a certain number of window slides, the error probability of nodes at position $z$ after $T$ iterations equals the initial error probability of nodes at position $z-1$ within the same window, for all $z \in \mathbb{Z}$. Furthermore, using the potential function method, we provide an upper bound on the constant wave propagation speed \( v = 1/T \), which characterizes the convergence speed of the WD.

\section{preliminaries}
\label{pre}
\subsection{LDPC Ensemble}

An undirected bipartite graph $G=(V \cup C, E)$ is defined as two disjoint sets of vertices $V$ and $C$, and a set of edges $E$, where $E$ is a subset of the pairs $\{\{v,c\}:v \in V, c \in C\}$.

\begin{figure}[htbp]
\centering
\subfigure[parity check matrix $H$]{
\raisebox{0.65\height}{
\includegraphics[width=1.3in]{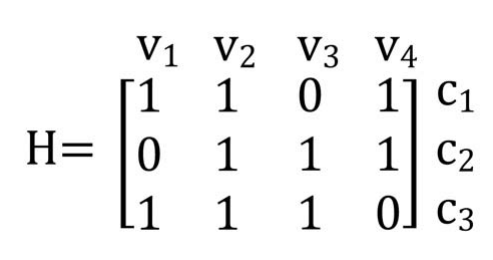}
}
\hfill
}
\subfigure[Tanner graph $G$]{
\includegraphics[width=1.3in]{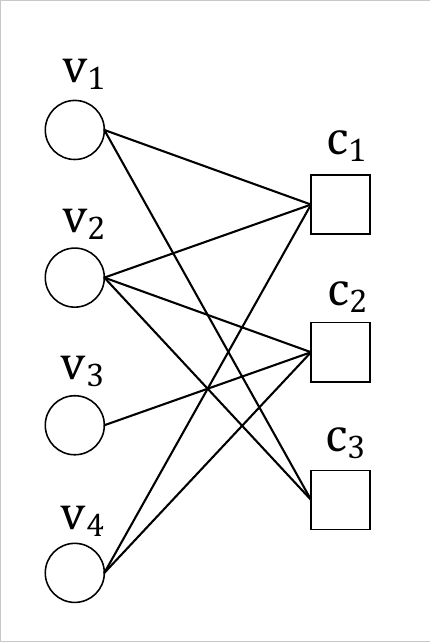}
}
\centering
\caption{Parity check matrix and Tanner graph for an LDPC code with a code length of $4$. In the Tanner graph, circles represent variable nodes, and squares represent check nodes. }
\label{fig1}
\end{figure}

An LDPC code is defined by a sparse binary parity-check matrix \( H \in \mathbb{F}_2^{m \times n} \), where \( n \) denotes the number of codeword bits and \( m \) denotes the number of parity-check constraints. The matrix \( H \) can be naturally represented as an undirected bipartite graph \( G = (V \cup C, E) \), known as the Tanner graph. In this graph, the nodes in \( V \), referred to as variable nodes, correspond to codeword bits, and the nodes in \( C \), referred to as check nodes, represent the parity-check constraints. An edge connects a variable node \( v_i \in V \) and a check node \( c_j \in C \) if and only if \( H(j, i) = 1 \). The parity-check matrix and Tanner graph for an LDPC code with a code length of $4$ are given in Fig. \ref{fig1}.

Let \(
L(x) = \sum_{i=1}^\infty L_i x^i \quad \text{and} \quad R(x) = \sum_{j=1}^\infty R_j x^j
\) be two polynomials, where \( L_i, R_j \in [0,1] \) for all \( i, j \), and \( \sum_{i=1}^\infty L_i = \sum_{j=1}^\infty R_j = 1 \). An LDPC code ensemble, denoted as \( \text{LDPC}(n, L, R) \), is defined as the set of all Tanner graphs with \( n \) variable nodes, the fraction of variable nodes of degree \( i \) is \( L_i \), and the fraction of check nodes of degree \( j \) is \( R_j \). And \( L(x) \) and \( R(x) \) are referred to as the degree distributions from the node perspective. For more details, see \cite{richardson2008modern}.

DE is a standard analytical technique used to evaluate the average bit error rate of the belief propagation algorithm over an LDPC code ensemble. Let \( \epsilon \) denote the erasure probability of the channel, and let \( x^{(t)} \) denote the average erasure probability of messages passed from variable nodes to check nodes at the \( t \)-th decoding iteration. We can write the DE equation as follows:
	\begin{equation}
	\label{DE}
		x^{(t)} = \epsilon \lambda(1-\rho (1-x^{(t-1)}))
	\end{equation}
where $\lambda(x) = L^\prime(x)/L^\prime(1)$ and $\rho(x) = R^\prime (x) / R^\prime(1)$. We set $x^{(0)} = 1$.
    
\subsection{Spatially Coupled LDPC Ensemble}
Let \( L(x) \) and \( R(x) \) denote the degree distributions. We now describe how to generate a random instance from the spatially coupled LDPC ensemble, denoted as \( \text{SC-LDPC}(L, R, N, w) \). There are \( M \) variable nodes at each position \( i \in [N] \triangleq \{1, 2, \dots, N\} \), and \( \frac{L'(1)}{R'(1)} M \) check nodes at each position \( j \in [N + w - 1] \). The degree of each variable (resp. check) node is randomly sampled according to the distribution \( L(x) \) (resp. \( R(x) \)). For a node of degree \( d \), we consider \( d \) sockets from which the $d$ edges emanate. To form the Tanner graph, for each socket of a variable node located at position \( i \), we first choose a position uniformly at random from the set \( \{i, i+1, \dots, i+w-1\} \), then randomly choose a socket from the check side at this position, and connect these two sockets to form an edge. After this, the remaining unconnected sockets are assumed to be connected to virtual variable nodes with zero erasure probability. We refer to \( N \) as the coupling length and \( w \) as the coupling width. For more details, refer to \cite{kudekar2011threshold}.

	
	
	\subsection{Window Decoding}
 WD is a decoding scheme designed based on the structure of spatially coupled LDPC codes, which allows for balancing performance with reduced complexity and decoding latency \cite{iyengar2011windowed}. Assume that the window size is \( W \). The first window consists of the variable nodes and check nodes from positions \( 1 \) to \( W \). The WD performs \( T \) iterations of BP decoding over the subgraph contained in the first window. Then, the window slides one position to the right, meaning the new window consists of the variable nodes and check nodes from positions \( 2 \) to \( W+1 \). Another \( T \) iterations are performed on the subgraph contained in the new window. This procedure proceeds until $T$ iterations have been completed on the subgraph consisting of the final $W$ positions of variable nodes and check nodes. During the decoding process, the window is referred to as the \( c \)th window configuration if it consists of the variable nodes and check nodes at positions \( c, c+1, \ldots, c+W-1 \) with $c \in [N-W+1]$.

Consider the SC-LDPC$(L, R, N, w)$ ensemble, let $x_z^{(c,t)}$ be the average erasure probability of a message incoming to check nodes in position $z \in \mathbb{Z}$ in the $c$th window configuration at iteration $t$. Let $\boldsymbol{x^{(c,t-1)}} = \{x_1^{(c,t-1)},\dots,x_{N+w-1}^{(c,t-1)}\}$, the DE equation \cite{iyengar2012windowed} is
\begin{equation}
\label{WD_DE1}
{x_z^{(c,t)}} =f(z,\boldsymbol{x^{(c,t-1)}};c,w,W,\epsilon),
\end{equation}
\begin{equation}
		\label{WD_DE2}
        \begin{split}
          &x_z^{(c+1,0)} = x_z^{(c,T)}, \\
          &\quad 
        \end{split}
	\end{equation}
 where $f(z,\boldsymbol{x^{(c,t-1)}};c,w,W,\epsilon)$	is defined as
	\begin{equation}
	\label{f_z}
		\begin{cases}
			\frac{1}{w} \sum\limits_{k=0}^{w-1} \epsilon_{z-k} \lambda(1-\frac{1}{w}\sum\limits_{j=0}^{w-1}\rho (1-x_{z+j-k}^{(c,t-1)}))        \\
			{\text{\;\;\;\;\;\;\;\;\;\ if}}\ z \in \{c,c+1,\cdots,c+W-1\} \\ 
			{x_z^{(c,t-1)},}       {\text{\;\;\;\;\;\;\;\;\;\; otherwise,}} 
		\end{cases}
		\end{equation}
\[\lambda(x) = L^\prime(x)/L^\prime(1),\rho(x) = R^\prime (x) / R^\prime(1),\]
for all $z \in [N+w-1]$, $c \in [N-W+1]$ and $t \in [T]$. We set $\epsilon_z=\epsilon$ for $z \in [N]$, and zero otherwise. The initial state of the decoding is set as $x_z^{(1,0)} = 1$ for all $z \in [N+w-1]$. For the boundary values, $z \notin [1,N+w-1]$, we set $x_z^{(c,t)} = 0$ for all $c \in [N-W+1]$ and $t\in \{0,1,\dots,T\}$. 
	
	\subsection{Potential Function}
The potential function defined on LDPC$(n, L, R)$ ensemble has been introduced and studied in \cite{aref2013convergence} and \cite{yedla2012simple1}. It is formally defined as follows: 
	\begin{equation}
		\label{U_LDPC}
		\begin{split}
			U(x;\epsilon) = \frac{1}{R^\prime(1)}(1-R(1&-x))-x\rho(1-x)\\
			&-\frac{\epsilon}{L^\prime(1)}L(1-\rho(1-x)).
		\end{split}
	\end{equation}

	Now consider the SC-LDPC($L, R, N, w$) ensemble. Similar to \cite{aref2013convergence}, the potential function under the $c$th window configuration is defined as follows:    
    \begin{equation}
	\label{WD_Udef}
	\begin{split}
		\frac{\partial}{\partial x_z} U(\boldsymbol{x};\epsilon,c,w,W) = \rho^\prime(1-x_z)\left(x_z-f(z,\boldsymbol{x};c,w,W,\epsilon)\right),
	\end{split}
	\end{equation}
where $\boldsymbol{x}=\{x_1,
\dots,x_{N+w-1}\}$ for all $z \in [N+w-1]$. It is straightforward to verify that the stationary points of the potential function correspond to the fixed points of the DE equation (\ref{WD_DE1}).
	
	Combining (\ref{f_z}), we can express the potential function under $c$th window configuration as follows:
    \begin{equation}
	\begin{split}	&U(\boldsymbol{x};\epsilon,c,w,W) =\sum_{z = c-(w-1)}^{c+W-1}  \frac{1}{R^\prime(1)}(1-R(1-x_z))-\\ &x_z\rho(1-x_z)-\frac{\epsilon}{L^\prime(1)}L\left(1-\frac{1}{w}\sum_{j=0}^{w-1}\rho(1-x_{z+j})\right).
	\end{split}
	\end{equation}
where $x_z = 0$ for all $z \notin [N+w-1]$. For notational simplicity, we omit $w$ and $W$ from the expression and write $U(\boldsymbol{x};\epsilon,c)$, with the understanding that $w$ and $W$ are fixed unless otherwise stated.
	
	Define the vector $\boldsymbol{y}$ as $y_z = x_z^{(c,t)} + h(x_z^{(c,t+1)}-x_z^{(c,t)})$ for $0 \le h \le 1$. Treat $\boldsymbol{x}^{(c,t)}$ as a known vector, we write the Taylor expansion of $U(\boldsymbol{y};\epsilon,c)$ around $U(\boldsymbol{x}^{(c,t)};\epsilon,c)$:
	\begin{equation}
		\begin{split}
			U(\boldsymbol{y};\epsilon,c) = &U(\boldsymbol{x}^{(c,t)};\epsilon,c) + \Delta U_1(\boldsymbol{y},\boldsymbol{x}^{(c,t)};c) \\
            &+ R_1(\boldsymbol{y},\boldsymbol{x}^{(c,t)};c)\\
            &\quad 
		\end{split}
	\end{equation}
	where $\Delta U_1(\boldsymbol{y},\boldsymbol{x}^{(c,t)};c) = \sum_{z=c}^{c+W-1} \frac{\partial U(\boldsymbol{y};\epsilon,c)}{\partial y_z} \Big|_{y_z = x_z^{(c,t)}} (y_z - x_z^{(c,t)})$, $R_1(\boldsymbol{y},\boldsymbol{x}^{(c,t)};c)$ is the remainder term.
	Citing the conclusion from \cite{aref2013convergence}, we have 
	\begin{equation}
	\label{daoshu}
	\alpha (U(\boldsymbol{y};\epsilon,c) - U(\boldsymbol{x}^{(c,t)};\epsilon,c)) \le \Delta U_1(\boldsymbol{y},\boldsymbol{x}^{(c,t)};c)
	\end{equation}
	with $1\le \alpha \le 2$. Although the results in \cite{aref2013convergence} pertain to BP decoding, a similar proof can be applied to reach this conclusion. Following \cite{aref2013convergence}, we set $\alpha =1$ in the subsequent simulation experiments.
	
	\begin{figure}
		\centering
		\includegraphics[width=0.6\linewidth]{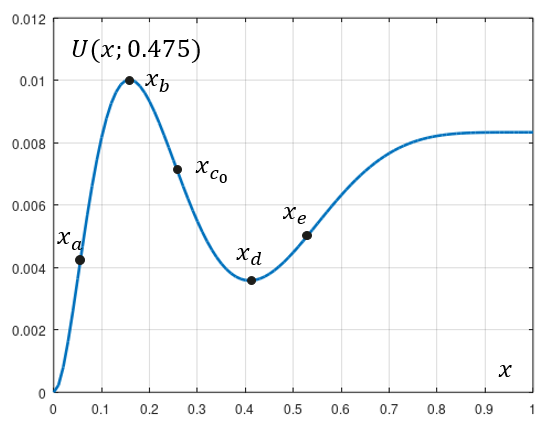}
		\caption{The potential function of the $(3,6)$-regular LDPC code ensemble for $\epsilon=0.475$. The stable $x_d$ and unstable $x_b$ fixed points of the DE equation are illustrated in the plot. In addition, $x_b$ and $x_{d}$ denote the $x$-coordinates corresponding to $|\frac{\partial U(x;\epsilon)}{\partial x}|=0$. $x_a,x_{c_0}$ and $x_e$ denote the $x$-coordinates corresponding to $|\frac{\partial^2 U(x;\epsilon)}{\partial x^2}|=0$.}
		\label{fig:1}
	\end{figure}
	
\section{Bounds on the convergence speed}

Kudekar, Aref, and others have studied the convergence speed of spatially coupled LDPC ensembles under conventional full BP decoding \cite{kudekar2012wave,aref2013convergence}. Specifically, it was proven that when the channel parameter is worse than the MAP threshold, information propagates in a wave-like manner with a constant propagation speed by progressing DE. However, this does not imply that the information propagates at a constant speed within a single window configuration under windowed decoding. Note that when the window has just slid to include nodes in positions \( c, c+1, \ldots, c+W-1 \), the messages associated with variable nodes at positions \( c, c+1, \ldots, c+W-2 \) have already been updated multiple times, whereas the messages at position \( c+W-1 \) remain at their initial values. This differs from the case of conventional full BP decoding, where all variable nodes are updated equally before wave-like information propagates, and leads to behavioral discrepancies between a single window configuration and conventional full BP decoding.




\subsection{The Phenomenon of Information Propagation Under WD}
For the SC-LDPC(\(L, R, N, w\)) ensemble, assume that there exists a constant \(c^\prime\) such that, for all \(c \ge c^\prime\), the DE solution \( \boldsymbol{x}^{(c,t)} \) satisfies
\begin{equation}
\label{assume}
x_{z+1}^{(c,t)} \ge x_z^{(c,t)} = x_{z+1}^{(c+1,t)}
\end{equation}
for all \( t \in \{0, 1, \dots, T\} \), and \( z \in [N + w] \). This assumption is supported by the fact that \( x_z^{(1,0)} = 1 \ge x_{z+1}^{(2,0)} \), and the monotonicity of (\ref{WD_DE1}) implies \( x_{z+1}^{(1,t')} \ge x_z^{(1,t')} \ge x_{z+1}^{(2,t')} \) for all \( t' \in [T]\) and \( z \) \cite{kudekar2011threshold}. By repeating this procedure and using (\ref{WD_DE2}), one obtains the inequality \( x_{z+1}^{(c,t)} \ge x_z^{(c,t)} \ge x_{z+1}^{(c+1,t)} \) for all \( c, z, t \). On the other hand, since the values \( x_z^{(c,t)} \) are bounded from below, this indicates that the sequence \( \{ x_{z^\prime+c}^{(c,t)} \}_{c} \) converges for all \( z^\prime \in \{0,1,\dots, W-2\} \) and \( t \in \{0,1,\dots, T\} \), which ensures the existence of such a constant \( c^\prime \).

\begin{figure}
		\centering
		\includegraphics[width=0.6\linewidth]{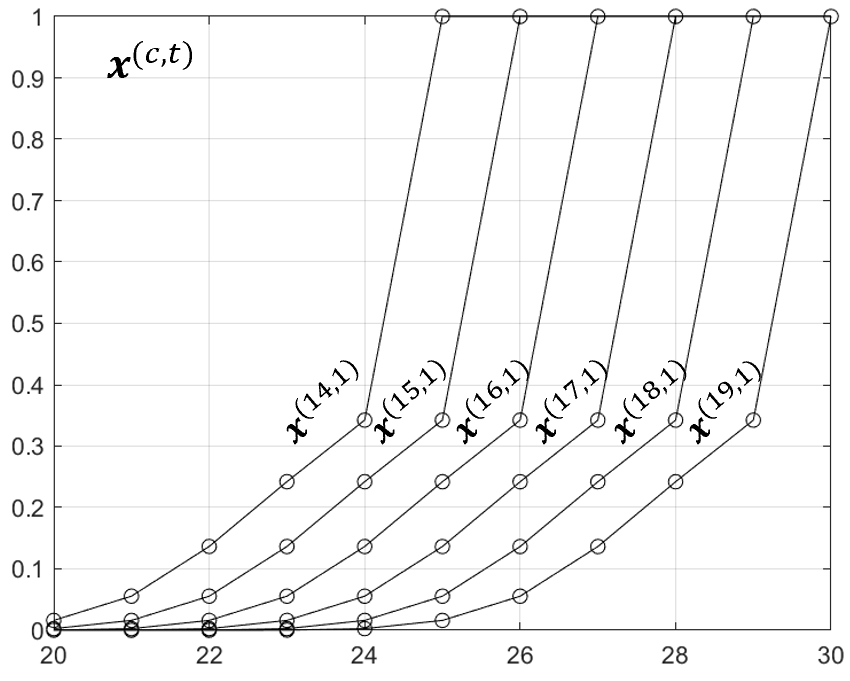}
		\caption{The DE solution for the ensemble SC-LDPC$(x^3,x^6,100,3)$ on a BEC with \(\epsilon = 0.42\) under different window configurations. The window size is 11, and each window configuration performs 6 iterations. }
		\label{fig:wd_1}
	\end{figure}

Experimental simulations also support the validity of the above assumption. For example, in Fig. \ref{fig:wd_1}, we plot the DE solution \( \boldsymbol{x}^{(c,1)} \) for the SC-LDPC\((x^3, x^6, 100, 3)\) ensemble with \( c \in \{14,15,\ldots,19\} \). It can be observed that \( x_z^{(c,1)} = x_{z+1}^{(c+1,1)},
 \) for all \( c \ge 14 \) and \( z \in [N+w] \).  

	\subsection{Bounds on the convergence speed under WD}
		\label{bounds_b}
Equation (\ref{assume}) shows that the DE solution \( \boldsymbol{x}^{(c,t)} \), which exhibits a wave-shaped profile, shifts one position to the right after \( T \) iterations, resulting in \( \boldsymbol{x}^{(c+1,t)} \). Define $v = 1/T$ to represent the propagation speed of the wave profile, which characterizes both the convergence speed of window decoding and the number of iterations required for successful decoding within each window. Next, we provide an upper bound on the propagation speed \( v \).	

		\begin{theorem}
		\label{th1}
 Consider the SC-LDPC$(L, R, N,w)$ ensemble and assume $T>1$. Let \( c^\prime \) be the constant such that 
 \begin{equation}
\label{assume_th1}
x_{z+1}^{(c,t)} \ge x_z^{(c,t)} = x_{z+1}^{(c+1,t)}
\end{equation}
holds for all \( c \ge c^\prime, t\in \{0,1,\dots,T\}\), and $z \in [N+w]$. Then,  
			\begin{equation}
				v \le A_1 :=\frac{\alpha \left(U(\boldsymbol{x}^{(c^\prime,0)};\epsilon,c^\prime) -U(\boldsymbol{x}^{(c^\prime+1,0)};\epsilon,c^\prime)\right)}{\sum_{z=c^\prime}^{c^\prime+W-1}\rho^\prime(1-x_z^{(c^\prime,0)})(x_z^{(c^\prime,0)}-x_{z-1}^{(c^\prime,0)})^2}
			\end{equation}
		\end{theorem}
The proof is provided in Appendix A. We can also provide an upper bound for \( v \), which eliminates the necessity of running DE.

Consider the SC-LDPC\((L, R, N, w)\) ensemble, where the DE equation (\ref{DE}) of the underlying ensemble LDPC\((L, R, n)\) has three fixed points: 0, an unstable fixed point \(x_b\), and a stable fixed point \(x_d\). Let \( U(x; \epsilon) \) be the potential function defined on the LDPC\((L, R, n)\) ensemble, and let \( x_a \) and \( x_{c_0} \) be two points satisfying \( \frac{\partial^2 U(x; \epsilon)}{\partial x^2} = 0 \), with \( x_a < x_{c_0} \).
		\begin{theorem}
	\label{th2}
Consider the SC-LDPC$(L, R, N,w)$ ensemble. Assume $T>1$ 
 and $x_{c-1}^{(c,0)} =0$ for all $c>1$. Let $D=\max_{x\in(0,x_d)}\left|\frac{\partial^2 U(x;\epsilon)}{\partial x^2}\right|$. Then
     \begin{equation}
	 	v \le \frac{w\alpha \left(U(1;\epsilon)\right)}{B_1},
	 \end{equation}
	 where
     
	\begin{equation}
	\begin{split}
		B_1 &= 2U(x_b;\epsilon)-U(x_d;\epsilon) + \frac{W\left(U^\prime(x_a;\epsilon)\right)^2}{D}\\
        &+ \frac{W\left(U^\prime(x_{c_0};\epsilon)\right)^2}{D}-D\frac{x_d}{w}.
	\end{split}
	\end{equation}
where $U^\prime(x;\epsilon) \triangleq \frac{\partial U(x;\epsilon)}{\partial x}$, and this simplifies for $w \to \infty$ to 
	$$
		v \le \frac{w\alpha \left(U(1;\epsilon)-U(e;\epsilon)\right)}{B_2},
	$$
	where $B_2$ is equal to
	\begin{equation}
	\begin{split}
 2U(x_b;\epsilon)-U(x_d;\epsilon) + \frac{W(U^\prime(x_a;\epsilon))^2 }{D}+ \frac{W(U^\prime(x_{c_0};\epsilon))^2}{D}.
	\end{split}
	\end{equation}
\end{theorem}
A brief proof of Theorem \ref{th2} is provided in Appendix B, while a more detailed derivation can be found in the supplementary material following the references. Although the analysis here focuses on the case where the window shifts by one position at a time, the same approach can be extended to cases where the window shifts by more than one position.

\section{SIMULATION RESULT}
\label{simu}
\begin{figure}
	\centering
	\includegraphics[width=0.6\linewidth]{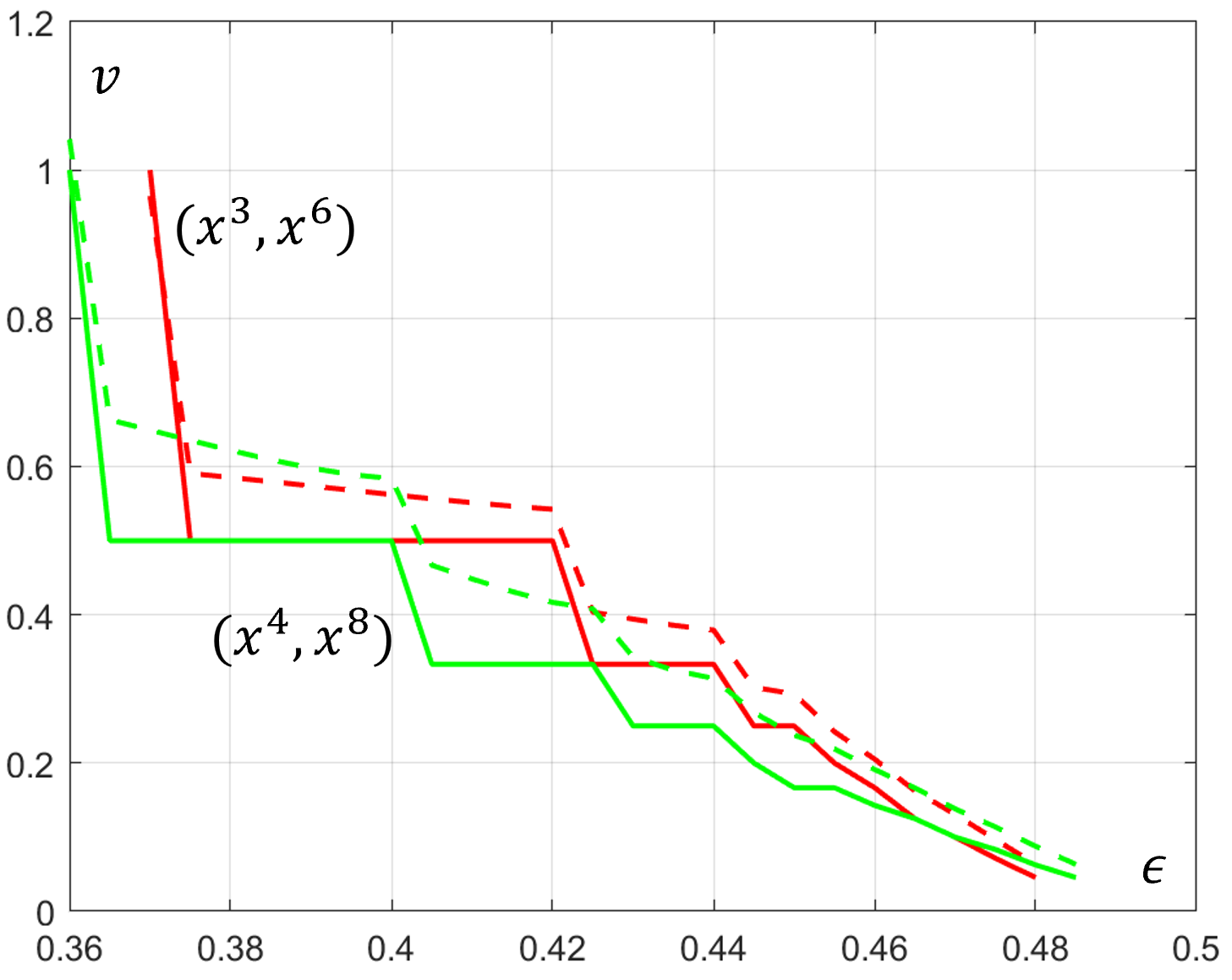}
	\caption{The plot of \( v \) vs. \( \epsilon \) for some SC-LDPC(\( x^k, x^{2k}, N, w = 4 \)) ensembles, with \( \epsilon \) less than the MAP threshold and a window size of 15 in each case. The upper bound \( A_1 \) is represented by dashed lines, while the information propagation speed calculated through DE is shown by solid lines.}
	\label{fig:re1}
\end{figure}

Through experiments, it was observed that the bound is tighter when $\alpha$ is set to 1. We therefore set $\alpha=1$ in the sequel. Consider the asymptotically regular SC-LPDC($x^k,x^{2k},N,w=4$) ensemble. In Fig. \ref{fig:re1}, we plot the propagation speed \( v \) that results in successful decoding (the average error probability below $10^{-6}$), along with its upper bound, for \( \epsilon \) below the MAP threshold with \( k = 3 \) and \( 4 \). The curves of the upper bound \( A_1 \) are shown as dashed lines, while the solid lines represent the propagation speed \( v \) computed via the DE equations. Since \( T=\frac{1}{v} \) is an integer, it is relatively insensitive to variations in \( \epsilon \), which leads to the staircase-like shape of the curve. It is observed that the propagation speed $v$ is a decreased function of $\epsilon$ and becomes zero at the MAP threshold. Moreover, it is also suggested that, under the same channel condition, the convergence speed can be improved by modifying the degree distribution.

We give $v$ and $A_1$ for $\epsilon=0.465$ and the different sizes of the window in Table \ref{table1}. We observe that the propagation speed increases with the increasing of the window size. When the window size gets larger, each variable node is included in more window configurations, allowing for more frequent updates compared to the case with a smaller window size. This results in improved performance and faster convergence. On the other hand, as the window size increases, the behavior of WD gradually approaches that of conventional full BP decoding, leading to a saturation in the convergence speed. Fig. \ref{fig:re1} and Table \ref{table1} illustrate how the propagation speed \( v \) varies with respect to the channel parameter \( \epsilon \) and the window size. The propagation speed effectively reflects the convergence behavior of WD and provides practical guidance on how to set the number of iterations in a single window.

\begin{table}[!t]
	\caption{propagation of SC-LPDC($x^3,x^{6},N,w=4$) ensemble with different size of window on BEC with  $\epsilon=0.465$}
	\centering
	\label{table1}
	\begin{tabular}{|c||c| |c| |c| |c|}
		\hline
		window size & 12 & 14 & 16 & 18\\
		\hline
		$v$ & 0.11 & 0.125 & 0.1429 & 0.1429\\
		\hline
		$A_1$ & 0.15 & 0.1624 & 0.1753 & 0.1753\\
		\hline
	\end{tabular}
\end{table}

\section{CONCLUSION}
\label{con}
This paper primarily investigates the convergence speed of spatially coupled LDPC ensembles under WD. Specifically, an upper bound on the convergence speed is derived, which can inform the design of the number of iterations required within each decoding window. Simulation results further demonstrate that, under the same channel conditions, ensembles with different degree distributions exhibit varying convergence speeds. This observation suggests the potential for future work aimed at designing degree distributions that optimize convergence speed.

\section*{Acknowledgment}
This work is partially supported by the National Key R\&D Program of China, (2023YFA1009600).

\appendices
\section{PROOF OF THE THEOREM \ref{th1}}
	Recall that $x_z^{(c,t)}$ is an increasing sequence in terms of $z$ and a decreasing sequence in terms of $t$. And \( c^\prime \) be the constant such that 
 \begin{equation}
\label{assume_appe}
 x_z^{(c,0)} = x_{z+1}^{(c+1,0)} = x_{z+1}^{(c,T)}
\end{equation}
holds for all \( c \ge c^\prime\) and $z \in [N+w]$. 
	According to (\ref{daoshu}), 
\begin{equation}
\begin{split}
\label{cw_xianglin}
	&\alpha \left(U(\boldsymbol{x}^{(c,t+1)};\epsilon,c) - U(\boldsymbol{x}^{(c,t)};\epsilon,c) \right)\\
	&\le \sum_{z=c}^{c+W-1}-\rho^\prime(1-x_z^{(c,t)})(x_z^{(c,t+1)}-x_z^{(c,t)})^2\\
    &\quad
\end{split}
\end{equation}
for $t \in \{0,1,\dots,T-1\}$. By summing (\ref{cw_xianglin}) with respect to $t$, we obtain:
\[
\begin{split}
		\label{T}
		&\alpha \left(U(\boldsymbol{x}^{(c,T)};\epsilon,c) - U(\boldsymbol{x}^{(c,0)};\epsilon,c) \right)\\
		&\le \sum_{t=0}^{T-1}\sum_{z=c}^{c+W-1}-\rho^\prime(1-x_z^{(c,t)})(x_z^{(c,t+1)}-x_z^{(c,t)})^2.
	\end{split}
\]
Now, consider the right-hand side of the above inequality and let us exchange the order of the sums. Then for each $z$, we minimize the following summation:
$$
\sum_{t=0}^{T-1} \rho^\prime(1-x_z^{(c,t)})(x_z^{(c,t+1)}-x_z^{(c,t)})^2 = \sum_{t=0}^{T-1} \rho^\prime(1-x_z^{(c,t)})l_t^2,
$$
where $l_t = x_z^{(c,t)}-x_z^{(c,t+1)}$. 

Due to (\ref{assume_appe}) and $x_z^{(c,t)}$ is a decreasing sequence in terms of $t$, $l_t>0$ and
$$
\sum_{t=0}^{T-1}l_t = x_z^{(c,0)} - x_z^{(c,T)} = x_z^{(c,0)} - x_{z-1}^{(c,0)}.
$$

The application of Jensen's inequality leads to 
$$
\frac{1}{T}\sum_{t=0}^{T-1}l_t^2 \ge \left(\frac{1}{T}\sum_{t=0}^{T-1}l_t\right)^2 = \frac{1}{T^2}(x_z^{(c,0)} - x_{z-1}^{(c,0)})^2,
$$
and thus,
\[
\begin{split}
	&\alpha \left(U(\boldsymbol{x}^{(c,0)};\epsilon,c) -U(\boldsymbol{x}^{(c+1,0)};\epsilon,c)\right)\\
	& \ge \sum_{z=c}^{c+W-1} \frac{1}{T} \rho^\prime(1-x_z^{(c,0)}) (x_z^{(c,0)} - x_{z-1}^{(c,0)})^2.
\end{split}
\]

\section{PROOF OF THE THEOREM \ref{th2}}

\section{supplementary material}
This is a more detailed version of the proof of Theorem \ref{th2}. In order to prove the theorem, we need to introduce an important lemma. The proof of the lemma is very similar to that in \cite{aref2013convergence} and is omitted here due to space constraints.  

\begin{lemma}
\label{cw_lemma_b1}
Let $x_z^{(c,t)}$ be the solution of DE equations (\ref{WD_DE1}) and (\ref{WD_DE2}) at some iteration $t$. Further, assume that $T_1>1$. Then,
$$
x_z^{(c,t)} - x_{z-1}^{(c,t)} \ge \left|\frac{x_z^{(c,t)}-\epsilon \lambda(1-\rho(1-x_z^{(c,t)}))}{w}\right|. 
$$
\end{lemma}

We now turn to the proof of Theorem \ref{th2}. To derive an upper bound on the convergence speed, we must lower-bound the following summation:
\begin{equation}
\label{cw_appB_1}
\sum_{z=c}^{c+W-1}\rho^\prime(1-x_z^{(c,0)})(x_z^{(c,0)}-x_{z-1}^{(c,0)})^2
\end{equation}

let $y_z = x_z^{(c,0)}$ for $z \in \mathbb{Z}$. From Lemma \ref{cw_lemma_b1}, we have
$$
\rho^\prime (1-y_z)(y_z-y_{z-1}) \ge \left| \frac{U^\prime (y_z;\epsilon)}{w} \right| \triangleq \left| \frac{\partial U(y_z;\epsilon)}{w\partial y_z} \right|.
$$
Thus,
$$\sum_{z=c}^{c+W-1}\rho^\prime(1-y_z)(y_z-y_{z-1})^2 \ge \sum_{z=c}^{c+W-1}\left| \frac{U^\prime (y_z;\epsilon)}{w} \right|(y_z-y_{z-1}). $$
\begin{figure}
	\centering
	\includegraphics[width=0.6\linewidth]{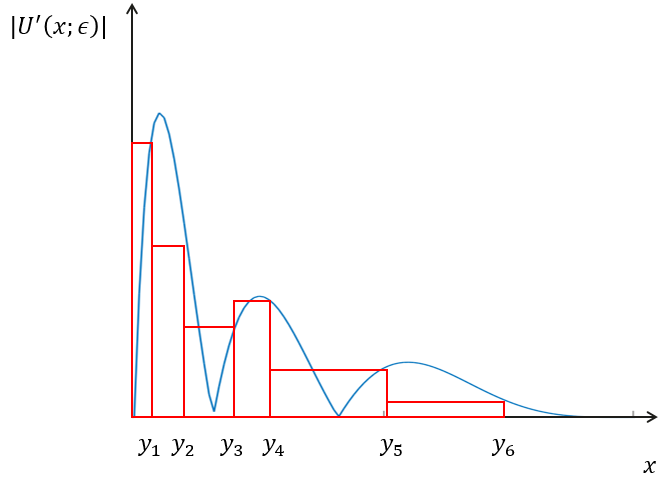}
	\caption{The absolute value of the derivative of the potential function of the $(3,6)$-regular LDPC code ensemble for $\epsilon=0.475$. The area of the red rectangle is $\sum_{z=1}^{6}\left| U^\prime (y_z;\epsilon) \right|(y_z-y_{z-1})$.}
	\label{cw_fig:b2}
\end{figure}
As shown in Fig.\ref{cw_fig:b2}, the lower bound of (\ref{cw_appB_1}) corresponds to the area of the red rectangle. We can divide the interval $[0,x_d]$ into four sub-intervals:
	\begin{enumerate}
		\item {$ x \in [0,x_a]$ where $U^\prime(x;\epsilon) \ge 0$ and $U^{\prime\prime}(x;\epsilon)\ge 0$.}
		\item {$ x \in [x_a,x_b]$ where $U^\prime(x;\epsilon) \ge 0$ and $U^{\prime\prime}(x;\epsilon)\ge 0$.}
		\item {$ x \in [x_b,x_{c_0}]$ where $U^\prime(x;\epsilon) \le 0$ and $U^{\prime\prime}(x;\epsilon)\le 0$.}
		\item {$ x \in [x_{c_0},x_d]$ where $U^\prime(x;\epsilon) \le 0$ and $U^{\prime\prime}(x;\epsilon)\le 0$.}
	\end{enumerate}
Let $a^\prime$ denote the index where $y_{a^\prime} \le x_a < y_{{a^\prime}+1}$. Similarly, we define $b,c_0$ and $d$ for $x_b,x_{c_0}$ and $x_d$, respectively.

(1) Since the potential functions becomes $\cap$-convex for $0 \le x \le x_a$, and the window size is $W$. Thus,
\begin{equation}
\label{cw_appB_2}
\begin{split}
&\sum_{z=c}^{{a^\prime}}\left| U^\prime (y_z;\epsilon) \right|(y_z-y_{z-1})+|U^\prime(x_a;\epsilon)|(x_a-y_{a^\prime})  \\
&= \sum_{i=1}^{W}\left|U^\prime (g_i;\epsilon) \right|(g_i-g_{i-1}),   
\end{split}
\end{equation}
where $g_i=y_{c+i-1}$ with $i=\{0,\dots,a^\prime-c+1\}$, $g_j=x_a$ with $j = \{a^\prime-c+2,\dots,W\}$. 

\begin{figure}
	\centering
	\includegraphics[width=0.9\linewidth]{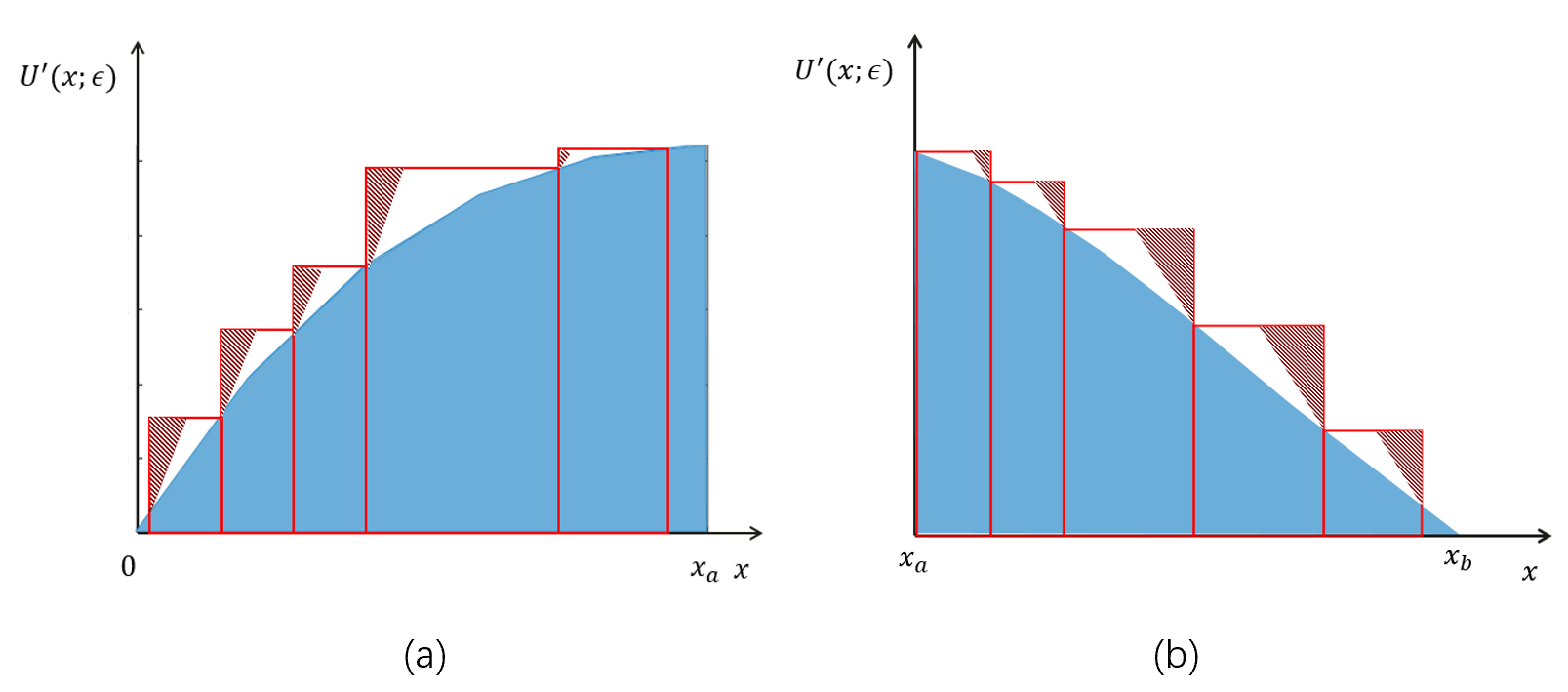}
	\caption{The derivative of the potential function of the $(3,6)$-regular LDPC code ensemble for $\epsilon=0.475$ over $x \in [0,x_a]$ for (a), and $x \in  [x_a,x_b]$ for (b). The area of the red rectangle represents $\sum_{z=c}^{a}\left|U^\prime (y_z;\epsilon) \right|(y_z-y_{z-1})$ and is composed of three parts: the blue area is the integral of the $U^\prime(x;\epsilon)$, the deep red area of the triangle, and the remaining white part.}
	\label{cw_fig:b1}
\end{figure}

 Let $D = max_{x \in [0,x_d]}|U^{\prime\prime}(x;\epsilon)|$, then 
\[
\begin{split}
&\sum_{i=1}^{W}\left|U^\prime (g_i;\epsilon) \right|(g_i-g_{i-1})  \\
&\ge \sum_{i=1}^W\bigg(U(g_i;\epsilon)-U(g_{i-1};\epsilon)\bigg)\\
&+ \sum_{i=1}^{W}\frac{1}{2D}\left(U^\prime(g_{i};\epsilon)-U^\prime(g_{i-1};\epsilon)\right)^2
\end{split}
\]
where $U(g_i;\epsilon)-U(g_{i-1};\epsilon)$ is the integral of the $U^\prime(x;\epsilon)$ over its domain $x \in [g_{i-1},g_i]$, $\frac{1}{2D}\left(U^\prime(g_{i};\epsilon)-U^\prime(g_{i-1};\epsilon)\right)^2$ is the area of a triangle with height $|U^\prime(g_i;\epsilon) - U^\prime(g_{i-1};\epsilon)|$ and slope equal to $D$. The integration region and the triangular area are illustrated in Fig.\ref{cw_fig:b1}. Since these two regions do not overlap, it follows that the sum of these two areas is upper bounded by $\left| U^\prime (g_i;\epsilon) \right|(g_i-g_{i-1})$.


The term $\sum_{i=1}^{W}\frac{1}{2D}\left(U^\prime(g_{i};\epsilon)-U^\prime(g_{i-1};\epsilon)\right)^2$ achieves the minimum value of 
$$
	 \frac{1}{2D}W (\frac{U^\prime(x_a;\epsilon)}{W})^2
$$
if and only if 
$$U^\prime(g_{i};\epsilon)-U^\prime(g_{i-1};\epsilon)=\frac{U^\prime(x_a;\epsilon)}{W}$$
for any $i \in \{1,\dots,W\}$. Thus,
\[
\begin{split}
&\sum_{z=c}^{{a^\prime}}\left| U^\prime (y_z;\epsilon) \right|(y_z-y_{z-1}) + \left| U^\prime (x_a;\epsilon) \right|(x_a-y_{a^\prime}) \\ &\ge U(x_a;\epsilon) + \frac{1}{2D}W (\frac{U^\prime(x_a;\epsilon)}{W})^2.
\end{split}
\]

(2) Since the potential functions becomes $\cap$-convex for $a^\prime+1\le z\le b^\prime$ and the mean value theorem, $\exists k_z\in[y_{z-1},y_z]$ such that
$$
U^\prime (y_z;\epsilon) - U^\prime (y_{z-1};\epsilon) = U^{\prime\prime} (k_{z};\epsilon)(y_z-y_{z-1}).
$$
Then,
\[
\begin{split}
&U^\prime(y_{a^\prime+1};\epsilon)(y_{a^\prime+1}-y_{a^\prime}) \\
&\ge(U^\prime(x_a;\epsilon)-D(y_{a^\prime+1}-x_a))(y_{a^\prime+1}-x_a+x_a-y_{a^\prime}) \\
&\ge U^\prime(x_a;\epsilon)(x_a-y_{a^\prime}) + U^\prime(x_a;\epsilon)(y_{a^\prime+1}-x_a)\\
&\;\;\;\;\;-D(y_{a^\prime+1} -y_{a^\prime} )^2,
\end{split}
\]
and 
$$
U^\prime(y_z;\epsilon)(y_z-y_{z-1}) \ge U^\prime(y_{z-1};\epsilon)(y_z-y_{z-1})-D(y_z-y_{z-1})^2,
$$
for all $a^\prime+2\le z\le b^\prime$. This leads to the conclusion that
\[
\begin{split}
&\sum_{z={a^\prime+1}}^{{b^\prime}}\left| U^\prime (y_z;\epsilon) \right|(y_z-y_{z-1})\\
   &=\sum_{z={a^\prime+1}}^{{b^\prime}}\left| U^\prime (y_z;\epsilon) \right|(y_z-y_{z-1}) +|U^\prime(x_b;\epsilon)|(x_b-y_{b^\prime})\\
   &\ge U^\prime(x_a;\epsilon)(x_a-y_{a^\prime}) + U^\prime(x_a;\epsilon)(y_{a^\prime+1}-x_a)\\
&\;\;\;\;+ \sum_{z={a^\prime+2}}^{b^\prime}U^\prime(y_{z-1};\epsilon)(y_z-y_{z-1}) -D\sum_{z=a^\prime+1}^{b^\prime}(y_z-y_{z-1})^2\\
&\;\;\;\;+U^\prime(y_{b^\prime};\epsilon)(x_b-y_{b^\prime})-D(x_b-y_{b^\prime})^2.
\end{split}
\]

Similar to $(1)$, we can regard $U^\prime(x_a;\epsilon)(y_{a^\prime+1}-x_a)+\sum_{z=a^\prime+2}^{b^\prime} U^\prime (y_{z-1};\epsilon)(y_z-y_{z-1})+U'(y_{b^\prime};\epsilon)(x_b-y_{b^\prime})$ as the area composed of three parts: the integral of the $U^\prime(x;\epsilon)$ over its domain $x \in [x_a,x_b]$, the area of the triangle, and the remaining part. Then 
\[
\begin{split}
&U^\prime(x_a;\epsilon)(y_{a^\prime+1}-x_a)+\sum_{z={a^\prime}+2}^{b^\prime} U^\prime (y_{z-1};\epsilon)(y_z-y_{z-1})\\
& +U'(y_{b^\prime};\epsilon)(x_b-y_{b^\prime}) \\
&\ge U(x_b;\epsilon) - U(x_a;\epsilon) + \frac{1}{2D}W (\frac{U^\prime(x_a;\epsilon)}{W})^2.
\end{split}
\]

According to the DE equation,
\[
\begin{split}
	&y_{z}-y_{z-1} = x_z^{(c,0)} -x_{z-1}^{(c,0)} \\
	& = \frac{1}{w}\Bigg(\epsilon_z \lambda(\frac{1}{w}\sum_{j=0}^{w-1}1-\rho(1-x_{z+j}^{(c,t-1)}))\\
	& -\epsilon_{z-w}\lambda(\frac{1}{w}\sum_{j=0}^{w-1}1-\rho(1-x_{z+j-w}^{(c,t-1)}))\Bigg)\\
	& \le \epsilon_z \lambda(\frac{1}{w}\sum_{j=0}^{w-1}1-\rho(1-x_{z+j}^{(c,t-1)})) \le \frac{\epsilon_z}{w}
\end{split}
\]

and 
\[
\begin{split}
	&\sum_{z=a^\prime+1}^{b^\prime} D(y_z-y_{z-1})^2+D(x_b-y_{b^\prime})^2 \\
    &\le \sum_{z=a^\prime+1}^{b^\prime} D\frac{\epsilon_z}{w} (y_z-y_{z-1})+D\frac{\epsilon_{b^\prime+1}}{w} (x_{b}-y_{b^\prime})\\
	&\le D\frac{x_{b}-y_{a^\prime}}{w}.
\end{split}
\]
Thus,
\[
\begin{split}
   &\sum_{z={a^\prime+1}}^{{b^\prime}}\left| U^\prime (y_z;\epsilon) \right|(y_z-y_{z-1}) \\
   &\ge U^\prime(x_a;\epsilon)(x_a-y_{a^\prime})+U(x_b;\epsilon) - U(x_a;\epsilon) \\
   &+ \frac{1}{2D}W (\frac{U^\prime(x_a;\epsilon)}{W})^2-D\frac{x_{b}-y_{a^\prime}}{w}.
\end{split}
\]

The case of $(3)$ is similar to that of $(1)$, and the case of $(4)$ is similar to that of $(2)$. These are omitted here due to space constraints.

By summing up all terms,
\[
\begin{split}
&\sum_{z=c}^{c+W-1}|U^\prime(y_z;\epsilon)|(y_z-y_{z-1}) \\
& \ge 2U(x_b;\epsilon)-U(x_d;\epsilon) + \frac{WU^\prime(x_a;\epsilon)^2}{D}+ \frac{WU^\prime(x_{c_0};\epsilon)^2}{D}\\
& -\frac{D}{w}(x_b-y_{a^\prime}+x_d-y_{{c_0^\prime}})\\
& \ge 2U(x_b;\epsilon)-U(x_d;\epsilon) + \frac{WU^\prime(x_a;\epsilon)^2}{D}+ \frac{WU^\prime(x_{c_0};\epsilon)^2}{D}\\
& -D\frac{x_d}{w}.
\end{split}
\]
And we have the result by noting that
\[
\begin{split}
  &\sum_{z=c}^{c+W-1}\rho^\prime(1-x_z^{(c,0)})(x_z^{(c,0)}-x_{z-1}^{(c,0)})^2 \\
  &\ge \sum_{z=c}^{c+W-1}\left| \frac{U^\prime (y_z;\epsilon)}{w} \right|(y_z-y_{z-1}).  
\end{split}
\]

\bibliographystyle{ieeetr}
\bibliography{ref}




\end{document}